Does the h$_\alpha$ index reinforce the Matthew effect in science?

Agent-based simulations using Stata and R

Lutz Bornmann*, Christian Ganser+, Alexander Tekles*+, & Loet Leydesdorff$

*Division for Science and Innovation Studies

Administrative Headquarters of the Max Planck Society

Hofgartenstr. 8,

80539 Munich, Germany.

Email: bornmann@gv.mpg.de

+Ludwig-Maximilians-Universität Munich

Department of Sociology

Konradstr. 6

80801 Munich, Germany.

$University of Amsterdam

Amsterdam School of Communication Research (ASCoR)

PO Box 15793,

1001 NG Amsterdam, The Netherlands.

Email: loet@leydesdorff.net

**Abstract**

Recently, Hirsch (2019a) proposed a new variant of the h index called the $h_α$ index. He formulated as follows: "we define the $h_α$ index of a scientist as the number of papers in the h-core of the scientist (i.e. the set of papers that contribute to the h-index of the scientist) where this scientist is the α-author" (p. 673). The $h_α$ index was criticized by Leydesdorff, Bornmann, and Opthof (2019). One of their most important points is that the index reinforces the Matthew effect in science. We address this point in the current study using a recently developed Stata command (h_index) and R package (hindex), which can be used to simulate h index and $h_α$ index applications in research evaluation. The user can investigate under which conditions $h_α$ reinforces the Matthew effect. The results of our study confirm what Leydesdorff et al. (2019) expected: the $h_α$ index reinforces the Matthew effect. This effect can be intensified if strategic behavior of the publishing scientists and cumulative advantage effects are additionally considered in the simulation.

**Key words**

bibliometrics, h index, $h_α$ index, Matthew effect, agent-based simulation, bibliometrics-based heuristics

# 1  Introduction

The h index introduced by Hirsch (2005) is one of the most-popular bibliometric indicators worldwide. The paper by Hirsch (2005) has been cited more than 3500 times (date of search in Web of Science, WoS, Clarivate Analytics: March 2019). The h index has been adopted as one among other indicators in WoS and Scopus (Elsevier). In the bibliometrics literature, however, many critical points have been raised about it: for example, Waltman and van Eck (2012) argue that "for the purpose of measuring the overall scientific impact of a scientist (or some other unit of analysis), the h-index behaves in a counterintuitive way. In certain cases, the mechanism used by the h-index to aggregate publication and citation statistics into a single number leads to inconsistencies in the way in which scientists are ranked" (p. 406). Furthermore, the counting of papers with citation numbers ≥h has not been justified by Hirsch (2005); it is equally possible to count papers with citation numbers ≥$h^2$ or h/2 (Egghe, 2006a; Egghe, 2006b).

Since the introduction of the h index many variants have been proposed targeting one or several disadvantages of the h index. Bornmann, Mutz, Hug, and Daniel (2011) concluded on the basis of a meta-evaluation that most of these variants correlate highly: "depending on the model, the mean correlation coefficient varies between .8 and .9. This means that there is redundancy between most of the h index variants and the h index" (p. 346). Recently, Hirsch (2019a) himself proposed a new variant called the $h_\alpha$ index: "we define the $h_\alpha$ index of a scientist as the number of papers in the h-core of the scientist (i.e. the set of papers that contribute to the h-index of the scientist) where this scientist is the α-author" (p. 673). Hirsch (2019a) recommended to use the new index in combination with the h index. The author formulated as follows: "a high h index in conjunction with a high $h_\alpha/h$ ratio is a hallmark of scientific leadership" (p. 673).

The h$_α$ index was criticized by Leydesdorff et al. (2019). One of their most important critical points is that the index "adds the normative element of reinforcing the Matthew effect in science" (p. 1163). The Matthew effect was defined by Merton (1968) as follows: "the Matthew effect consists in the accruing of greater increments of recognition for particular scientific contributions to scientists of considerable repute and the withholding of such recognition from scientists who have not yet made their mark" (p. 58). Merton (1968) cites a physicist as follows: "The world is peculiar in this matter of how it gives credit. It tends to give the credit to (already) famous people" (p. 57). The Matthew effect is very similar to Price's (1976) "cumulative advantages" that he noted as a core mechanism in the sciences explaining, among other things, the skewed distributions in the indicator values. Barabási (2002) reinvented Price's cumulative advantages and Merton's Matthew effect as "preferential attachment" without any knowledge of or reference to this bibliometric literature (Bonitz, Bruckner, & Scharnhorst, 1999).

Hirsch (2019b) partly denied that the h$_α$ index reinforces the Matthew effect in science: "Strictly speaking at most half of this is true, the higher h-index author in a collaboration benefits, however the lower h-index author does not get negatively affected, his/her h$_α$ remains the same. More importantly, lower h-index authors have the choice to not collaborate with high h-index authors but rather pursue their own independent work, or work with more junior collaborators" (p. 1168).

We agree with Hirsch (2019b) that (probably) authors or co-authors with low h index values will not become "poorer" and nothing is taken away from them. The first problematic point in his statement is, however, the implicit demand to search strategically for cooperation in science. Following the norms in the ethos of science (Merton, 1942, 1973), cooperating partners should be selected based on the quality of their research or the fit to the needed expertise for a certain research project, but not for non-scientific reasons such as the increase of indicator values. Supervisor-supervised relationships might be other reasons for co-authorship besides h index values that counteract a co-author selection based on purely scientific reasons (however, which can scarcely be avoided in science).

Since Hirsch (2019b) partly rejected our claim, the second problematic point in his statement above is the remaining uncertainty about the reinforcement of the Matthew effect by using $h_\alpha$ in research evaluation. Thus, we address this point in the current study. We used a recently developed Stata command (h_index) and R package (hindex), which can be used to simulate h index and $h_\alpha$ index applications in research evaluation. Based on fictitious data the user can empirically investigate whether $h_\alpha$ reinforces the Matthew effect or not.

## 2        Literature overview and conceptual roots

### 2.1      The role of simulations in scientometrics

Albeit that simulations are not in the focus of the bibliometric literature, both bibliometrics and simulation studies have been used as quantitative methods in quantitative science and technology studies (e.g., Ahrweiler, 2001; Edmonds, Gilbert, Ahrweiler, & Scharnhorst, 2011; Scharnhorst, Börner, & van den Besselaar, 2012). Gilbert (1997) set the stage with the first simulations of the structure and dynamics of academic science. He introduced "kenes" as knowledge-variants of genes; the resulting events showed Lotka-type distributions and were interpretable using Simon's (1957) models of social processes. Ahrweiler – in collaboration with two co-authors – developed a large innovation model called SKIN: "Simulating Knowledge Dynamics in Innovation Networks" (Ahrweiler, Pyka, & Gilbert, 2004, 2011).

Different from data-oriented studies, simulations enable us to theorize mechanisms and to specify expectations. Not only observable behavior but also coordination and selection mechanisms can be studied. Leydesdorff and den Besselaar (1998), for example, showed that the Cobb-Douglas production function can be elaborated into a representation of technological trajectories and technological regimes by assuming feedback mechanisms (Leydesdorff & van den Besselaar, 1994). In a similar vein, one can simulate lock-ins and deadlocks in technological innovation (Leydesdorff, 2001; Leydesdorff & van den Besselaar, 1998) and synergy in Triple-Helix models (Ivanova & Leydesdorff, 2014). In the confrontation with data, the insights in mechanisms can be developed into what Bornmann and Marewski (in press) further elaborated into bibliometrics-based heuristics (BBH, see section 2.3).

During the early 2000s, this focus on the content of science and technology in more abstract (knowledge-based) terms disappeared because of the popularity of agent-based modeling in neighboring disciplines (Edmonds, Hernandez, & Troitzsch, 2007; Tesfatsion, 2002). Leydesdorff (2015) argued for a focus on (genotypic) mechanisms instead of phenotypical behavior. From this perspective, the observable dynamics of the sciences can be studied evolution-theoretically (Campbell, 1991; Distin, 2010; Hodgson & Knudsen, 2011; Ionescu & Chopard, 2013; Popper, 1972).

Meyer, Lorscheid, and Troitzsch (2009) provide a bibliometric analysis of the first decade of the *Journal of Artificial Societies and Social Simulations* (JASSS). The Matthews effect itself has extensively been simulated (for example, in physics) under the heading of preferential attachment (Abbasi, Hossain, & Leydesdorff, 2012; Barabási, 2002; Barabási et al., 2002; Bonitz et al., 1999; Garavaglia, van der Hofstad, & Woeginger, 2017; Newman, 2001; Petersen et al., 2014).

In a recent study, Backs, Günther, and Stummer (2019) used agent-based modeling as decision support system when planning measures to encourage academic patenting within universities. The authors suggest "the application of agent-based modeling and simulation, an approach that has been successfully used in other, similar, contexts (e.g., when selecting useful measures for market introduction and diffusion of new products). We have presented herein an agent-based model that is suitable for this purpose, and we have demonstrated its applicability and its potential value for practice (i.e., TTO [technology transfer offices] management drives increased patenting) and subsequently for society (i.e., more academic patents lead to an increase in knowledge transfer between universities and industry and/or provide a basis for spin-off companies) by means of an application example" (p. 454).

You, Han, and Hadzibeganovic (2016) used an agent-based simulation model to assess how the impact of scientists' work efficiency and their capability to select important topics for their research affects the h index (and other measures). In this simulation model, the agents (authors or research teams) try to occupy nodes in a citation network (publications). By providing the citation network a priori, the simulations focus on the process of competing for possible publications, rather than the collaboration or the citation process. The model proposed by You et al. (2016) is an example of how the influence of individuals' actions on macro-level patterns can be analyzed by means of simulations in scientometrics.

Besides that, we are aware of only a few simulation studies in scientometrics which focus on the h index. These simulations – as a rule – have dealt with the development of single h index values without considering collaborations between scientists. Lobet (2016) published an h index evolution simulator which reveals the development of single h index values based on various inputs (e.g., starting year of publishing, papers per year). The simulator is able to consider certain behaviors of researchers, for example, to always cite own papers. Guns and Rousseau (2009) investigated the h index's growth based on computer simulations of publication and citation processes. They found that "in most simulations the h-index grows linearly in time. Only occasionally does an S-shape occur, while in our simulations a concave increase is very rare" (p. 410). Ionescu and Chopard (2013) published two agent-based models which refer to performance measurements of single scientists and a group of scientists (see also Żogała-Siudem, Siudem, Cena, & Gagolewski, 2016). They studied, for example, what happens when low h-index researchers are removed from a community. Their results suggest "a stratified structure of the scientific community, in which the lower h levels mostly cite papers from the upper h levels" (p. 426).

## 2.2 Analytical sociology

This study follows the approach of analytical sociology which focusses on the mechanisms leading to social phenomena (Hedström, 2005; Hedström & Ylikoski, 2010). It is the goal of analytical sociology to work out the mechanisms (on the micro level) which are the causes of the phenomena (on the macro level) (Bornmann, 2010). In this study, we are interested whether the phenomenon "Matthew effect" can be produced by the mechanism "$h_\alpha$ index". In our simulation, the action is on the micro-level, since action (publishing, being cited, collaborating, and performance measuring) is done on the single-agent level, and the possible outcome is on the macro-level – structures in the form of certain $h_\alpha$ index distributions. In order to test the relationship between mechanism and phenomenon in this study, several agent-based simulations have been performed by using the Stata h_index command (and the corresponding R package). Most of the model parameters are held constant over all simulations; compared to a baseline simulation, only one parameter is changed in each of another three simulations to inspect the effect of this parameter. The interested reader of this paper can use the command or package to investigate the effect of further parameter variations.

## 2.3 Bibliometrics-based heuristics

The h_index command and the hindex package can be used to define rules for running various simulations. For example, we work with certain distributions of h index values as starting points and define how the agents in the simulation interact. The simulations are used then to receive an experimental view on the effects of the $h_\alpha$ index use in research evaluation.

Recently, Bornmann and Marewski (in press) introduced BBHs. They discuss the use of bibliometrics in research evaluation against the backdrop of the fast-and-frugal research program (e.g., Gigerenzer, Todd, & ABC Research Group, 1999) and define BBHs as decision strategies in research evaluation which ignore many, but use limited information (data) about an entity (i.e., citation and publication data of a researcher) to assess the entity. The application of heuristics in many other environments, for instance business, medicine, sports, and crime (Gigerenzer & Gaissmaier, 2011) has shown that they come to similar good judgments than more complex decision strategies.

By following the fast-and-frugal research program, Bornmann and Marewski (in press) define for the use of BBHs some search, stopping, and decision rules. These rules help to define and apply BBHs for a certain research evaluation environment. For example, these rules can be defined as follows: In economics, publications in so-called top-five journals (*American Economic Review*, *Econometrica*, *Journal of Political Economy*, *Quarterly Journal of Economics*, and *Review of Economic Studies*) decide about scientific careers (Bornmann, Butz, & Wohlrabe, 2018); reaching a professorship without having published in these journals is frequently not possible. The search, stopping, and decision rules for filling a professorship can be defined as follows: (1) search for all publications of a group of candidates (economists); (2) stop search when all publications have been identified; (3) select the candidate with the most papers in top-five journals.

One important element in the fast-and-frugal research program is the investigation of heuristics in certain environments: do they come to reliable and valid judgements? Is the application of certain bibliometric indicators in the environment reasonable? Does the indicator's use lead to non-desired effects? In this paper, we follow the BBH approach by studying the possible advantages and disadvantages of the $h_\alpha$ index use in research evaluation. We especially focus on the assumed sensitivity of the $h_\alpha$ index for the Matthew effect.

## 3 Implementation of our simulation model in Stata and R

The ado h_index and the hindex package simulate agents who collaborate on publishing papers. In Stata, type `net install h_index, from(https://raw.githubusercontent.com/chrgan/h_index/master/)` to install the ado.[1] The R package hindex can be installed by typing `devtools::install_github("atekles/hindex")`.[2] The simulation procedure is as follows:

(1) As a starting point, *n* agents are generated. The user can specify *n*, the number of agents. The agents have published in the past. The user can choose between a Poisson or negative binomial distribution for the number of previously published papers and set parameters of the distribution. It is assumed that each paper has been written one to five periods ago (imagine years, for example). For a share of these papers, which the user can specify interactively, the agent is the alpha author.

(2) For this set of *n* agents, the h index and h alpha index are calculated.

(3) Then, the agents start to collaborate. The previously published papers might be the result of collaborations among the simulated agents or with other agents. This does not matter for the rest of the simulation procedure. The user can specify how many periods the agents collaborate. In each period, the agents form teams publishing new papers. The user can set some properties: the average number of co-authors, the share of agents who collaborate in each period, and the correlation between the probability of co-authoring with other agents and the h index values calculated in step 2. Thus, one can specify that agents with high initial h index values are more productive than agents with low initial h index values. By default, the collaborating agents are assigned to co-authorships at random. However, it is possible to specify that agents with high h index values avoid co-authorships with agents who have equal or higher h index values. In this case, the agents with high h index values strategically select co-authors to improve their $h_\alpha$.

---

[1] The Stata module moremata has to be installed in advance (Jann, 2005).
[2] The package devtools has to be installed for this option to install the hindex package.

(4) All papers receive citations each period. The number of citations depends on (a) the citation distribution and (b) the age of the paper.

(a) The user can choose between a Poisson and a negative binomial distribution. He/she can specify the maximum expected number of citations.

(b) The expected number depends on the papers' age following a log-logistic function. It first increases with time periods, reaches the maximum specified in step 4(a) after a configurable number of periods and then decreases. The steepness of the log-logistic function can be specified, too.

Thus, for each given age of the papers, the number of citations follows the distribution specified in step 4(a) with an expected citation number given by its maximum and the age of the paper. A graph showing the distribution of the expected values can be generated.

To reflect the possibility of self-citations, the user can specify an option leading to one additional citation for each paper (published at least one period ago) where at least one of its authors has an h index value which exceeds the number of previous citations of the paper by one or two. This reflects agents strategically citing their own papers which have citations just below their h index value. This accelerates the growth of the agents' h index values. Finally, a "boost" effect can be specified: papers of agents with higher h index values are cited more frequently than papers of agents with lower h index values.

(5) For each period, the new values of h and h alpha are calculated. The alpha author of a paper can be determined at the time of its publication (without changing later on) or the alpha author of a paper is determined after each period of action based on the current h index values of the authors (see Tietze, Galam, & Hofmann, 2019).

(6) To ensure the robustness of the results, steps 1 to 5 are repeated *r* times.

# 4   Results

The Matthew effect implies that the more reputable scientist receives more credit than the less reputable scientist for a scientific contribution, although the contribution is of the same scientific quality. Thus, the credit is not attributed fairly on the basis of the performed contribution, but (unfairly) on the basis of previous contributions. If we compare this definition of the Matthew effect with the definition of the $h_\alpha$ index, the similarities are obviously observable. In case of the $h_\alpha$ index, the credit for a paper is assigned to the co-author with the highest h index. Although all authors conributed to the co-authored paper in question, only one author receives the full credit. Furthermore, the credit is assigned to the co-author who is most reputable in terms of h index values. These similarities between the definitions of Matthew effect and $h_\alpha$ already point out that the simulations which are presented in the following can be expected to reveal the appearance of the Matthew effect by using the $h_\alpha$ index in performance measurement.

## 4.1   First agent-based simulation with 200 agents (baseline simulation)

Similar to the BBHs program with search, stopping, and decision rules (see above), the first agent-based simulation has three phases: initial setting, acting (collaborating) several periods, and final dataset for further analysis (visualization of the results). Whereas the initial setting and the final dataset is on the macro level (certain distributions are set or analyzed), acting is on the micro level (see section 2.2). It is the goal of the first agent-based simulation – the baseline simulation, compared to which one parameter is changed in each of the simulations presented in the following sections – to compare the mean $h_\alpha$ index values of agents with initial low or high h index values after several periods of action (e.g., collaboration with other agents). The Stata command for the first agent-based simulation is

```
h_index, r(50) n(200) per(20) co(3) dp(poisson, mean(10))
dc(poisson, mean(5)) p(3) sh(.33) clear
```
[3].

**Initial setting**: The first simulation is based on 200 agents [`n(200)`]. The agents in the groups have published on different output and impact levels: the distribution of the papers follows a Poisson distribution, the agents have published 10 papers on average [`dp(poisson, mean(10))`]. For 1/3 of all papers published by an agent, the agent itself is the alpha-author (-agent) [`sh(.33)`]. h index and $h_\alpha$ index values are calculated for all agents.

**Acting**: Agents act (publish, collaborate, receive citations) across 20 periods [`per(20)`]. Each collaborating group of agents has three agents on average [`co(3)`]. The citations, which the co-authored papers published by the agent groups receive, follow a Poisson distribution with a specified time-dependent expected value [`dc(poisson, mean(5))`]. The time-dependent expected value follows a log-logistic distribution reaching its maximal value of 5 after 3 years (following the general guideline by Glänzel & Schöpflin, 1995) [`p(3)`]. The agent-based simulation is repeated 50 times [`r(50)`] to ensure the robustness of the simulation. After each simulation, new h index and $h_\alpha$ index values are calculated for all agents.

---

[3] The equivalent function call to produce the simulated data in R is:
```
simulate_hindex(runs = 50, n = 200, periods = 20, coauthors = 3,
distr_initial_papers = 'poisson', dpapers_pois_lambda = 10, distr_citations
= 'poisson', dcitations_mean = 5, dcitations_peak = 3, alpha_share = .33)
```

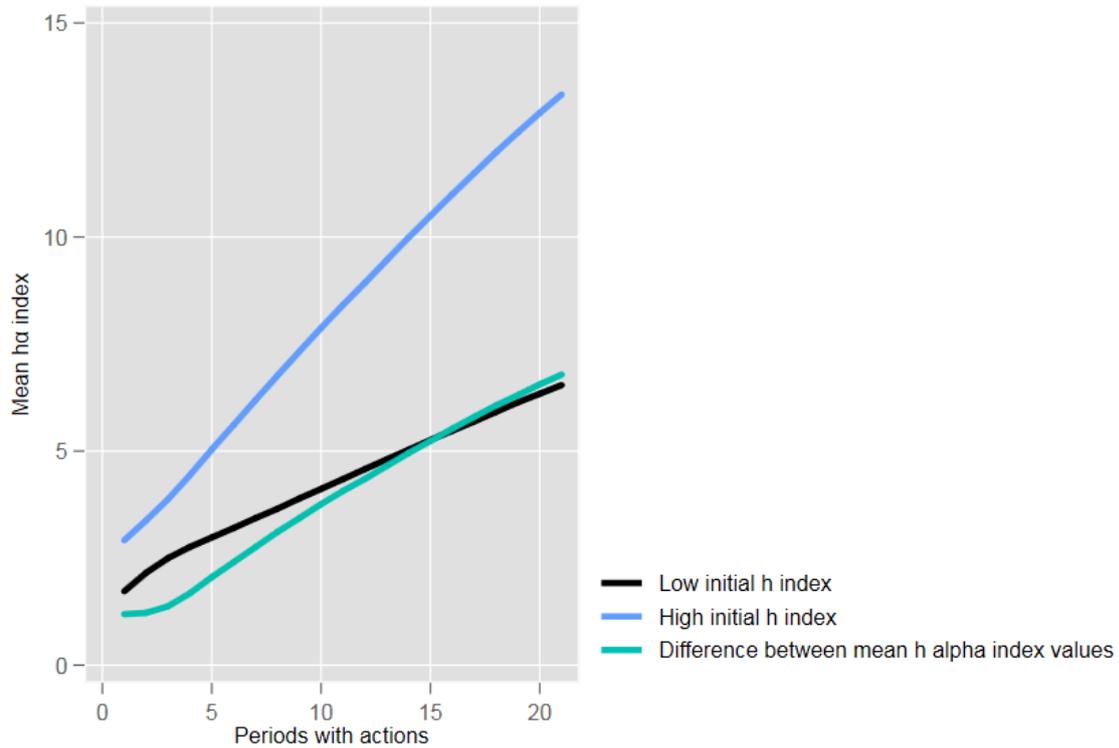

Figure 1. Results of the first agent-based simulation: mean $h_\alpha$ index values for two groups of agents with low (<7) and high (>7) initial h index values

**Final dataset**: Two groups of agents are refined with low (<7) and high (>7) initial h index values (7 is the median initial h index value). For each period with actions (20 in total), the mean $h_\alpha$ index values are computed (across 50 repetitions of the simulation to have robust results). The results are shown in Figure 1. For each period with actions, the advantage of the agents with high initial h index values is clearly visible: they do not only start with higher mean $h_\alpha$ index values (which is as expected), these values also increase with additional periods – by considering further cooperation, publications and additional citations. The mean $h_\alpha$ index values of the agents with low initial h index values also increase over time. However, the difference between the two groups becomes larger with onward periods – as the green line in Figure 1 demonstrates. Increasing differences between both groups can be interpreted as a Matthew effect in operation.

## 4.2 Second agent-based simulation with an additional element leading to more citations for prolific agents

The second simulation has been run using the Stata command `h_index, r(50) n(200) per(20) co(3) dp(poisson, mean(10)) dc(poisson, mean(5)) p(3) sh(.33)` **`boost(size(.5))`**[4]. It is the same command as in the first agent-based simulation (the baseline simulation), but we introduce a new element with `boost(size(.5))` (which is printed in boldface). This option means that papers published by agents with higher h index values are cited more frequently than papers published by agents with lower h index values. The results of Frandsen and Nicolaisen (2017) demonstrate, for instance, that "authors in the field of Healthcare tend to cite highly cited documents when they have a choice. This is more likely caused by differences related to quality than differences related to status of the publications cited" (p. 1278).

The number of citations in the second simulation are increased based on the value specified with [`size(.5)`]. For example, suppose agents with a maximal h index value of 11 have published a certain paper. The value .5 as option means that this paper receives round(11*.5)=6 additional citations.

**Final dataset**: In the second agent-based simulation, the median of the initial h index values (median=7) is the same as in the first simulation. Thus, two groups of agents are refined with low (<7) and high (>7) initial h index values. Figure 2 presents the results. The results are similar to Figure 1, but the differences between both groups are more pronounced: whereas the $h_\alpha$ index values of the group with high initial h index values increase more steeply, the $h_\alpha$ index values of the group with high initial h index values increase similar to those in Figure 1. This leads to larger mean $h_\alpha$ index values differences between both groups

---

[4] The equivalent function call to produce the simulated data in R is:
`simulate_hindex(runs = 50, n = 200, periods = 20, coauthors = 3, distr_initial_papers = 'poisson', dpapers_pois_lambda = 10, distr_citations = 'poisson', dcitations_mean = 5, dcitations_peak = 3, alpha_share = .33, boost = TRUE, boost_size = .5)` (as the green line reveals). In other words, the Matthew effect is reinforced by letting the papers published by agents with higher h index values be cited more frequently than the agents with lower h index values.

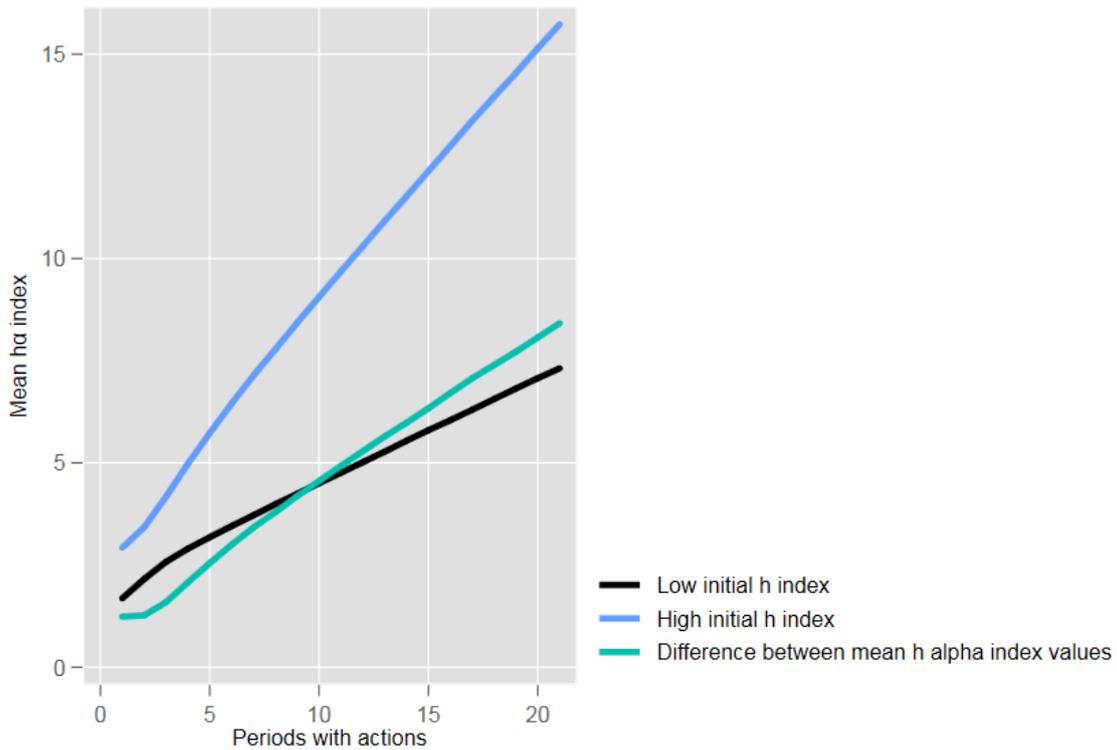

Figure 2. Results of the second agent-based simulation: mean $h_\alpha$ index values for two groups of agents with low (<7) and high (>7) initial h index values

### 4.3 Third agent-based simulation considering the correlation of new citations with h index values: agents with high h index values receive disproportional citations

For the third simulation, the following Stata command has been used: `h_index, r(50) n(200) per(20) co(3) dp(poisson, mean(10)) dc(poisson,`

`mean(5)) p(3) sh(.33) `**`dil(correlation(.8) share(.6))`**[5]. Similar to the second simulation, only one option has been changed (which is printed in boldface) in comparison to the first baseline simulation. The new options [`dil(correlation(.8) share(.6))`] focusses on the probability of publishing new papers depending on initial h index values. The option [`correlation(.8)`] means that agents with high initial h index values are more productive than agents with low initial h index values: the correlation between the probability of publishing new papers and initial h index values has been set to .8. The option [`share(.6)`] means that 60% of the agents publish. The use of this option can be reasoned, for instance, by the so called "sacred spark" theory (Cole & Cole, 1973) which claims "that there are substantial, predetermined differences among scientists in their ability and motivation to do creative scientific research" (Allison & Stewart, 1974, p. 596).

The third agent-based simulation is intended to check whether the higher productivity of prolific agents has an effect on the $h_\alpha$ index values development of the groups with high and low initial h index values.

---

[5] The equivalent function call to produce the simulated data in R is:
```
simulate_hindex(runs = 50, n = 200, periods = 20, coauthors = 3,
distr_initial_papers = 'poisson', dpapers_pois_lambda = 10, distr_citations
= 'poisson', dcitations_mean = 5, dcitations_peak = 3, alpha_share = .33,
diligence_corr = .8, diligence_share = .6)
```

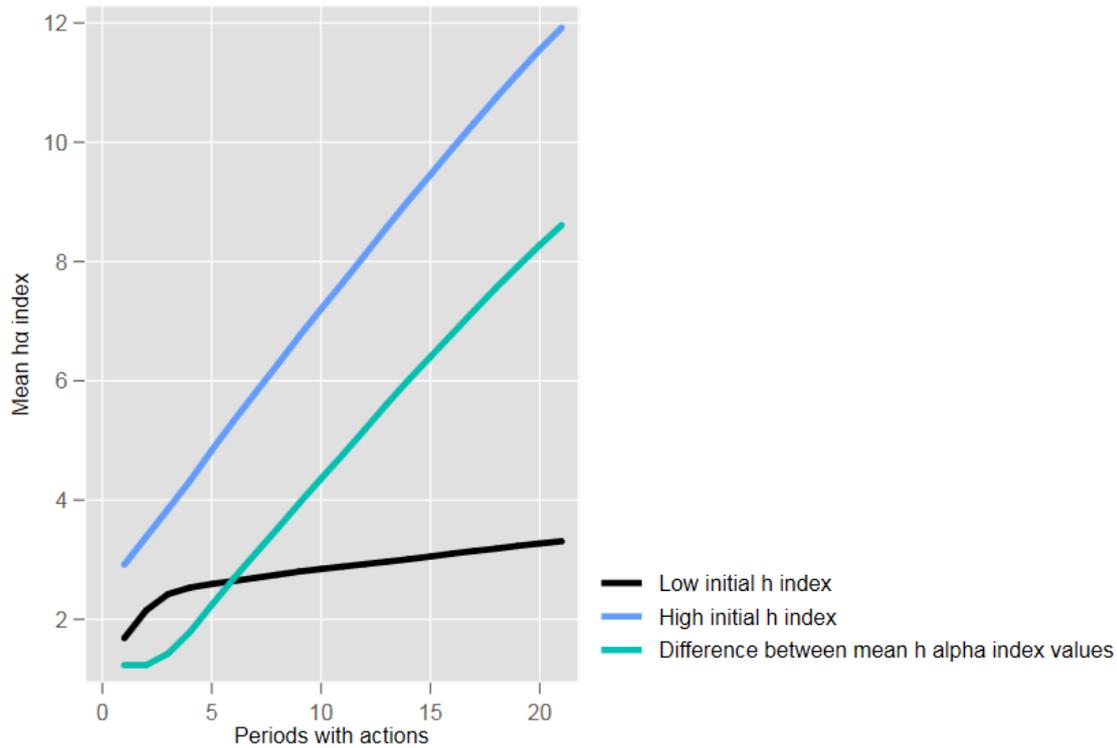

Figure 3. Results of the third agent-based simulation: mean $h_\alpha$ index values for two groups of agents with low (<7) and high (>7) initial h index values

**Final dataset**: The results of the third simulation are presented in Figure 3. Whereas this third simulation considers a positive correlation between productivity and h index values, the second simulation includes the positive relationship between citations and h index values (see Figure 2). The findings in Figure 3 vary significantly from the results in Figure 1 and Figure 2. As the green line for the differences between the mean $h_\alpha$ index values reveals, the differences increase stronger over time than in Figure 2. Thus, the results of both simulations point out that an effect on the $h_\alpha$ index values can be especially expected when additional output is included, while the effect is less pronounced if additional impact is modeled instead.

### 4.4  Fourth simulation considering strategically selecting co-authors

For the fourth simulation, we used the Stata command `h_index, r(50) n(200) per(20) co(3) dp(poisson, mean(10)) dc(poisson, mean(5)) p(3)`

`sh(.33) clear st`[6]. Compared to the baseline simulation, we additionally considered a strategic element [`st`], which focuses on the possible tendency of agents to select other agents as co-authors with lower h index values. Such a strategical element (with another focus) has been mentioned by Hirsch (2019b): "lower h-index authors have the choice to not collaborate with high h-index authors but rather pursue their own independent work, or work with more junior collaborators" (p. 1168). The `strategic` option of the h_index command means that a single agent with an h index – as high as possible – is assigned to every group of collaborating agents. Then, all other agents in the simulation are randomly allocated to the collaborating groups. Thus, the `strategic` option sizes the idea of collaborating with lower h index agents. The `strategic` option gives much weight to the effect of strategic collaboration decisions in our simulations, since the agents with the highest h index values never collaborate, so that their $h_\alpha$ index values increase after every collaboration. Even though this specification puts a lot of emphasis on the strategic component in the collaboration process, the results of this simulation reveal the potential effect of strategic collaboration decisions on the outcome distribution.

The strategic option follows closely the Coleman's (1990) classic macro-micro-macro model (i.e., "Coleman's boat"). "The general thrust of this model is that proper explanations of macro-level change and variation entail showing how macro-states at one point in time influence the behavior of individual actors, and how these actions add up to new macro-states at a later time" (Hedström & Swedberg, 1996, p. 296). The model assumes that individual action results from the social context in a social network. Coleman's model for the fourth agent-based simulation (see Figure 4) starts with the possible influence of a social context on the attitudes of agents (A). The current situation in science is characterized by performance based evaluations: "Especially in universities, government funding of scientific research is increasingly based upon performance criteria. As research institutions operate more and more in a global market, international comparisons of institutions are published on a regular basis" (Moed, 2018). This situation puts pressure on agents doing science in the system.

---

[6] The equivalent function call to produce the simulated data in R is:
```
simulate_hindex(runs = 50, n = 200, periods = 20, coauthors = 3,
distr_initial_papers = 'poisson', dpapers_pois_lambda = 10, distr_citations
= 'poisson', dcitations_mean = 5, dcitations_peak = 3, alpha_share = .33,
strategic_teams = TRUE)
```

The second (B) and third (C) steps are characterized by the core components of Hedström's (2005) DBO theory (desires, beliefs, and opportunities). (B) The second step in the macro-micro-macro model is that the social context (here: increasing focus on performance criteria) influences the attitudes of single agents: The agents **believe** (given the pressure in the system) that they should increase their $h_\alpha$ index values. As acting agents in the system they **desire** to perform as good as possible in terms of bibliometric indicators. (C) The agents have several **opportunities** to act: they can collaborate with other agents without considering their h index values or they can consider that in their reflections (among other alternatives). Since the $h_\alpha$ index can only be improved when agents publish papers with co-authors having lower h index values, the `strategic` option simulates this possible tendency of agents. (D) The empirical analyses of the development of $h_\alpha$ index values for agents with low and high initial h index values across several periods of action reveal how single actions of agents lead to the social phenomenon on the macro level: the reinforcement of the Matthew effect.

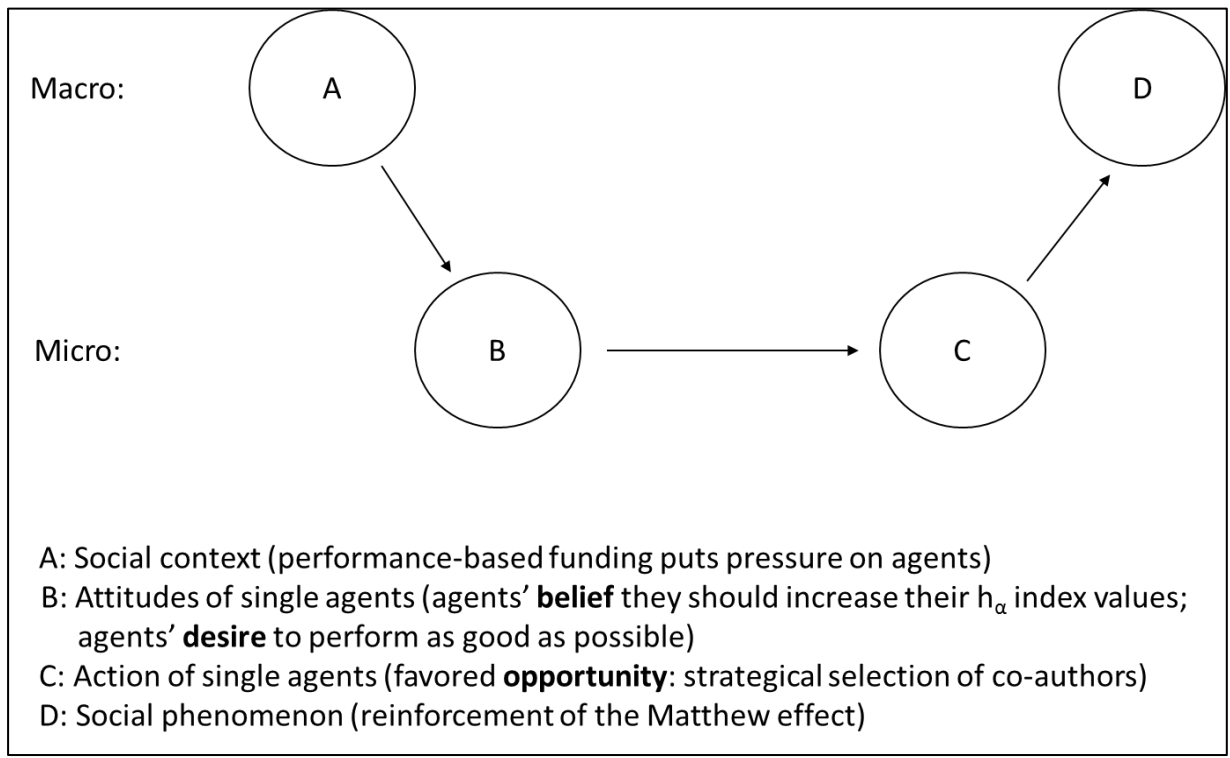

Figure 4. Coleman's (1990) macro-micro-macro model depicting the relationship between performance-based funding and reinforcement of the Matthew effect

The result of the fourth agent-based simulation is shown in Figure 5. It is clearly visible that the strategic element significantly reinforces the Matthew effect which is already visible in the previous simulations: agents with low h index values not only have lower initial $h_\alpha$ index values than agents with high h index values, the $h_\alpha$ index values also increase on a significantly lower level across the periods of evaluation. Across the periods of actions, the $h_\alpha$ index value differences between both h index groups become larger and larger.

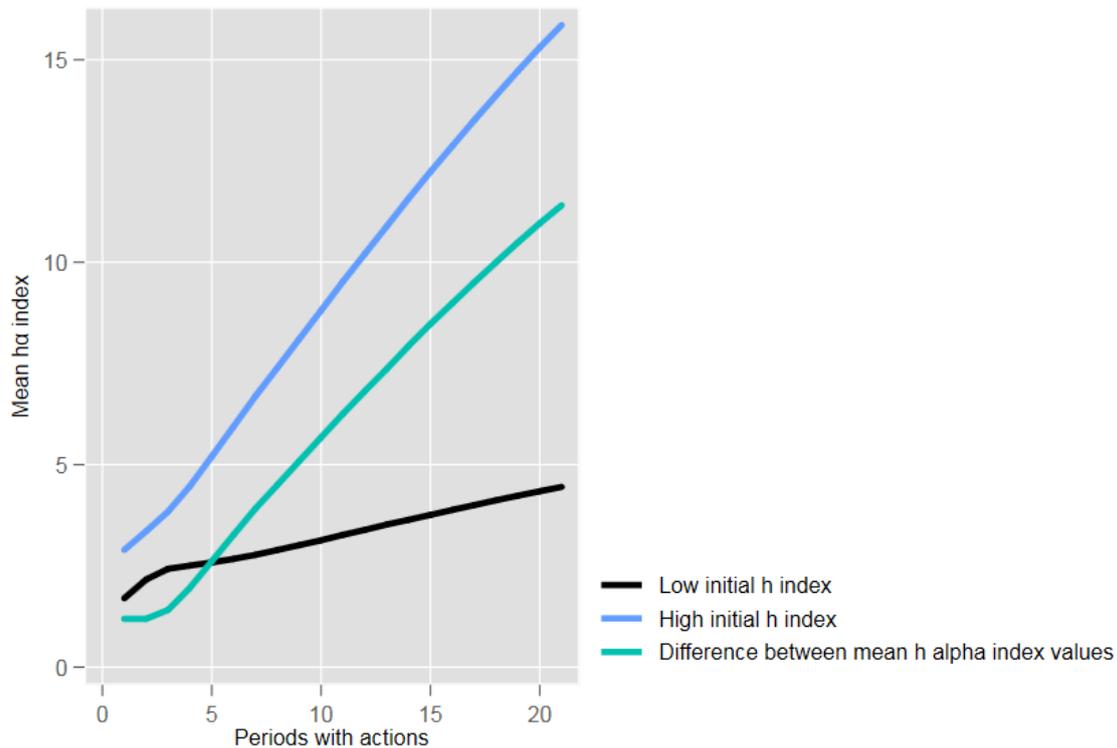

Figure 5. Results of the fourth simulation: mean $h_\alpha$ index values for two groups of agents with low (<7) and high (>7) initial h index values

## 5  Discussion

The agent-based simulations presented in this paper follow a recent discussion in *Scientometrics* about the newly introduced $h_\alpha$ index by Hirsch (2019a). Leydesdorff et al. (2019) assumed that the use of the new index reinforces the Matthew effect in research evaluations. Scientists with initial high h index values will profit disproportionally from the use of the $h_\alpha$ index. Thus, the fear is that the use of the index enlarges a problem, which is already prevalent in the science system. According to Merton (1968) the problem of the Matthew effect in science is so great that "we are tempted to turn again to the Scriptures to designate the status-enhancement and status-suppression components of the Matthew effect. We can describe it as 'the Ecclesiasticus component', from the familiar injunction 'Let us now praise famous men', in the non-canonical book of that name" (p. 58).

Leydesdorff et al. (2019) even assumed – based on the definition of the $h_\alpha$ index – that the disproportional attribution of credit by the $h_\alpha$ index – the co-author with the highest h index receives the full credit – reflects the operation of the Matthew effect. Thus, the $h_\alpha$ index is already the Matthew effect in operation. One cannot assume that the co-author with the highest h index contributes so much to the paper that the other co-authors can be completely discarded in performance measurement. In this study, we abstained from the single case and tested with various simulations whether the Matthew effect is visible on the macro level – when certain reasonable initial parameters and parameters for acting are set. The results of our study confirm what we expected from the single case: the $h_\alpha$ index reinforces the Matthew effect. This effect can be intensified if strategic behavior of the publishing scientists and accumulative advantage effects are additionally considered in the simulation.

Our study is situated in the tradition of the analytical sociology which seeks for mechanism-based explanations. These explanations try to focus on the crucial elements of a given process and to abstain from the detailed view (Hedström & Ylikoski, 2010). Agent-based modeling is a way of connecting the individual with the social level (Hedström, 2005). For studying a certain phenomenon on the macro level, the environment is defined in which the action takes place. Then, the action is running following certain pre-defined rules. It follows a dataset which constitutes the result of actions and initial parameters. This dataset can be used to investigate whether the social phenomenon of interest is observable on the macro level. By varying certain parameters of an agent-based model used as baseline, the effect of various situational elements from publishing, being cited, and collaborating on the development of the distribution of $h_\alpha$ index values can be tested.

This study is not only rooted in the analytical sociology, but also in the BBHs program. The program demands that indicators are empirically studied whether they can be used in certain evaluation environments (and if so, how they can be used). The h_index command and hindex package which we introduced in this paper can be used to simulate the use of the h index and $h_\alpha$ index in certain pre-defined environments. Using different specifications of the command (package functions), the simulation can be adapted to the environment for investigating where the $h_\alpha$ index is intended to be used. In this study, we used the Stata command to test whether the Matthew effect becomes apparent when the $h_\alpha$ index is calculated for a group of agents who collaborate, publish, receive citations across several periods. The command could be applied, for instance, by deans of faculties to decide whether the $h_\alpha$ index should be used for research evaluation or not. Is there the danger that the Matthew effect becomes apparent? Introducing some parameters (data) in the model which the dean investigates beforehand (how many scientists are in the faculty, how is the mean and distribution of output, etc.), the dean can study whether the Matthew effect emerges or not.

The R package and Stata command allow to consider some strategic elements in the agent-based simulations: if the $h_\alpha$ index is used in research evaluation processes, scientists might try to cooperate strategically with co-authors having lower h index values. The findings of our simulations reveal that the consideration of this element leads to a significant reinforcement of the Matthew effect. Using different options of the h_index command or different parameters for the hindex package functions, the agent-based simulations can not only consider strategic behavior, but also information from the literature on the usual behavior of scientists and distributions of publications and citations in different fields and institutions (e.g., Perianes-Rodrigueza & Ruiz-Castillo, 2014). For example, we considered in our agent-based simulations that agents with higher h index values will publish more frequently than agents with lower h index values. Many studies have shown that future performance depends on previous performance (Abramo, D'Angelo, & Soldatenkova, 2017; Allison, Long, & Krauze, 1982; Kwiek, 2015). We also included another element in the simulations which can be derived from the literature: that authors might tend to cite highly-cited papers.

Since the R package and Stata command are freely available, we encourage their use. We plan to add further functionality to them in the near future.

# Acknowledgement

We thank Jorge Hirsch for encouraging discussions and very useful comments to a preliminary version of this manuscript.